\documentclass[12pt,a4paper]{article}
\usepackage{amssymb} 
\usepackage[]{graphicx}
\begin{document}
\title{Consistent initial data for CMD perturbations}

\author{Wojciech Czaja$^2$, Zdzis{\l}aw A. Golda$^{1,2}$, Andrzej Woszczyna\thanks{uowoszcz@cyf-kr.edu.pl}\,\,\,$^{1,2}$}
\date{}
\maketitle

\vspace{-10mm}

\begin{center}
{{$^1$} Astronomical Observatory, Jagiellonian University,\\  ul. Orla 171, 30--244 Krak\'ow, Poland}\\
\smallskip
{{$^2$} Copernicus Center for Interdisciplinary Studies,\\ ul. S{\l}awkowska 17, 31--016 Krak\'ow, Poland}
\end{center}

\bigskip
\bigskip 
\arraycolsep.20em
%

\def \XX{{\eta,x}}
\def\KX{{k x}}
\def \ph{\phantom}
\def \pez{\phi(\XX)}
\def \pez{\phi}
\def \es{\eta_\de}
\def \pjez{\phi_{(1)}(\eta,x)}
\def \pdez{\phi_{(2)}(\eta,x)}
\def \aeta{a(\eta)}
\def \a2eta{a^2(\eta)}
\def \apeta{a'(\eta)}
\def \diag{\mathop{\rm diag}} 
\def \pp#1#2{\frac{\partial#1}{\partial#2}}
\def \pps#1#2{\frac{\partial^2#1}{\partial#2^2}}
\def \cc{\mbox{c.c.}}
\def \tk{\tilde{k}}
\def \te{\tilde\eta}
\def \tom{\tilde\omega}
\def \tti{{\tt i}}
\def \tz{\tilde{z}}
\def \mon{{}}
\def \amp{{\cal N}}
\def \ka{\tilde{k}}
\def \te{\tilde{\eta}}
\def \om{\tilde{\omega}}
\def \iks{\tilde{x}}
\def \ka{{\kappa}}
\def \te{{\tau}}
\def \om{{\omega}}
\def \iks{{\chi}}
\def\t1{{(1+\te)}}
\def\polowka{\left[\frac{1+\te}{2}\right]}
\def\tp125{{\left(\frac{\t1}{2}\right)^5}}
\def\todwr{{\left[\frac{2}{1+\te}\right]^5}}
\def\gr400{{400(4-4\ka^2+2\ka^4)}}
\def\gr800{{800(2-2\ka^2+\ka^4)}}
\def\jeden{{\left(1-3\ka^2 +4\ka^4\right)}}
\def\trzy{{(3-4\ka^2 +\ka^4)}}
\def\cztery{{\left(4+\ka^2 \t1^2\right)}}
\def\dziewiec{{\left(9+3\ka^2 +\ka^4\right)}}
\def\mian2{{2\left(-6+\ka^2\ \t1^2\right)}}
\def\ud{{\left[\frac{\mian2}{\tp125}\right]}}
\def\TCa{{ f}}
\def\TCb{{g}}
\def\Ta{2(\kappa^2(1+\tau)^2-6)\left(\frac{1+\tau}{2}\right)^{-5}}
\def\Tb{4+\kappa^2(1+\tau)^2}
\def\ik{{{\tt i} \ka}}
\def\ikate{{{\tt i} \ka \te}}
\def\2ikiks{{{\tt i} \ka \iks}}
\def\YX{{\te,\iks}}
\def \v4{{U}}
\def \de{{d}}
\def \Journal#1#2#3#4{{#1} {\bf #2}, #3 (#4)}
\def \AiA{Astron.~Astropys.}
\def\AJ{Astron.~J.}
\def \AP{Adv.~Phys.}
\def \ApJ{Astrophys.~J.}
\def \ApJL{Astrophys. J. Lett.}
\def \JETP{Sov.~Phys.~JETP}
\def \MNRAS{Mon.~ Not. R.~Astron. Soc.}
\def \ASS{Astroph. Space Sci.}
\def \CQG{Class. Quantum Grav.}
\def \JMP{J.~Math. Phys.}
\def \NPB{{Nucl. Phys.}~B}
\def \NC{{Nuovo Cimento}}
\def \PLA{{Phys. Lett.}~A}
\def \PR{Phys. Rep.}
\def \PRD{{Phys. Rev.}~D}
\def \PRd{{Phys. Rev.}}
\def \PRL{Phys. Rev. Lett.}

\begin{abstract}
We investigate the initial condition which are simultaneously consistent 
with perturbation equations 
in both the radiation-dominated 
and the mater-dominated epochs. 
The exact formula for the spectrum transfer-function is derived.
\end{abstract}

\section{Introduction\label{sec01}}

The scale invariance of the perturbations growth in the matter dominated epoch 
is known as the result of both the Newtonian and Einstein gravity theories. It is caused by the absence of the Euler term in the propagation equations~\cite{Peebles1980, Brandenberger&Kahn&Press1983}. Partial differential equations derived as the first order expansion of Einstein equations, immediately reduce to ordinary ones. 
According to standard procedures the initial conditions for these equations are being imposed at the beginning of the matter dominated epoch. Initial velocity field is set to zero throughout the space, or alternatively, the decaying mode is neglected~\cite{Zeldovich&Novikov}. The perturbation then is the product of time-dependent factor and an arbitrary function of the space coordinates. 
Perturbations of any shape and any scale grow equally slow~\cite{Peebles1980,Weinberg1972, Landau&Lifshitz1975} --- too slow to form compact objects in the time interval admitted by observations~\cite{Bouwens2010, Bunker2010, McLure2010}. 

Similar constraints for initial conditions are often applied to the $n$-body simulations~\cite{Bagla&Padmanabhan}. Following~\cite{Efstathiou1985} numerous treatments additionally adopt the Zeldovich approximation~\cite{Zeldovich1970} or its generalizations~\cite{Li&Barrow2010}. The perturbations statistics and amplitudes are confronted with the CMB data and inflationary scenarios~\cite{Nicol2007, Boyanovsky&Wu2010}. Most approaches adopt Newtonian gravity or the Newtonian limit of GR. The $n$-body simulations are promissory in non-linear regime. The question how to reach nonlinear regime still remains  difficult for both simulations and analytical solutions.

In this paper we investigate initial condition which are simultaneously consistent 
with  the perturbation equations in both, the radiation- and mater-dominated epochs.
Following the techniques elaborated in the gravitational waves theory~\cite{Grishchuk1974, AbbotHarari1986,Grishchuk1997, Allen2000, Durrer&Rinaldi2009, Baskaran2010}, we treat 
both epochs (with the perturbation corrections included) as the single Riemannian manifold. 
Appropriate junction conditions guarantee that the first and the second fundamental forms at the decoupling hypersurface $\Sigma_d$ are continuos. The perturbation modes are unique and regular in the entire spacetime. The {momentum spectrum} of sound in the radiation-dominated epoch (the gauge-invariant and time-independent Fourier coefficients $A_k$)
uniquely determine the perturbations evolution, both prior and past to decoupling.
Due to  junction conditions the perturbations in the late epoch ($p=0$) inherit some properties from the epochs before decoupling. One of the symptoms is the breakdown of the late-time scale-invariance.

\section{Brief description of the method}

Einstein equations and Darmois--Israel conditions
provide solution for the background scale factor
	\begin{eqnarray}
a_{(1)}(\eta)&=&\sqrt{\frac{{\cal M}}{3}}\,\eta,\\
a_{(2)}(\eta)&=&\frac14\sqrt{\frac{\cal M}{3}}\,\frac{(\eta+\eta_\de)^2}{\eta_\de}.
	\end{eqnarray}
$\eta_\de$ stands for the time of decoupling, ${\cal M}$ is a constant 
of motion, ${\cal M}=a\,\epsilon^4$ in the radiation filled universe.
In the conformal Newtonian gauge with the metric tensor 
\begin{eqnarray}
{\sf g}_{\mu\nu}=\a2eta\diag 
	\left(
-1-2\pez,\,1-2\pez,\,1-2\pez,\,1-2\pez
	\right).\nonumber\\
	\label{eq:metryka}
	\end{eqnarray}
the propagation equations for the density perturbations in
both epochs take forms
	\begin{eqnarray}
\frac{1}{\eta}\pp{}{\eta}\left[\frac{1}{\eta^2}\pp{}{\eta}\eta^3 \pjez\right]
-\frac{1}{3}\nabla_{\mu}\nabla^{\mu}\pjez=0,
	\label{eq:PertRad}
	\end{eqnarray}
for the radiation-dominated epoch, and
	\begin{eqnarray}
\pp{}{\eta}\left[\frac{1}{(\eta+\es)^4}\pp{}{\eta}(\eta+\es)^5 \pdez\right]=0.
	\label{eq:PertDust}
	\end{eqnarray}
for the matter-dominated one.

New perturbation variable $\Psi$ defined as 
the Darboux transform~\cite{Zwillinger} of the potential $\pjez$
	\begin{eqnarray}
\Psi{(\XX)}=\frac{1}{\eta}\pp{}{\eta} \left[\eta^3\,\pjez\right],
	\end{eqnarray}
allows to reduce (\ref{eq:PertRad}) to  d'Alem\-bert equation\footnote{For other gauges or gauge-invariant variables the equivalent procedures are given in~\cite{2001CQGra..18..543G, 2003PhLA..310..357G, 2005AcPPB..36.2133G, 2004A&A...419..801S}.}
	\begin{eqnarray}
\left[
\pps{}{\eta}-\frac{1}{3}\nabla_{\mu}\nabla^{\mu}
\right]
\Psi(\XX)=0.\label{DAl}
	\end{eqnarray}
$\Psi$ is an analogue of the massless scalar 
field. The general solution to (\ref{DAl}) is the linear combination of running waves
	\begin{eqnarray}
\Psi(\XX)=\sum_k {\cal A}_k{\sf u}_k(\XX)+\cc,
	\end{eqnarray}
where the modes 
	\begin{eqnarray}
{\sf u}_k(\XX)=\frac{1}{\sqrt{2\pi}}\exp\left[{{\tt i}(\KX-\omega\eta)}\right]
	\end{eqnarray}
are harmonic in $\eta$, and normalized
	\begin{eqnarray}
\langle{{\sf u}}_k(\XX),{{\sf u}}_{k'}(\XX)\rangle=\omega(k)\,\delta^3(k-k')
	\end{eqnarray} 
according to the Klein--Gordon scalar product
	\begin{eqnarray}
\langle
\Psi_1(\XX),
\Psi_2(\XX)
\rangle
={\tt i}\!\int\!\!\left[
{\Psi}_1\partial_\mu{\Psi}_2^*-{\Psi}_2\partial_\mu{\Psi}_1^*
\right]
d\sigma^\mu.\nonumber\\
\label{KG}
	\end{eqnarray}
The integral in formula (\ref{KG}) runs over arbitrary space-like hypersurface $\sigma^\mu$,
therefore, the coefficients $A_k$ are gauge-invariant and constant in time. 
Now, the field $\pez$ in the radiation-dominated epoch is a combination
	\begin{eqnarray}
\phi_{(1)}(\XX)=\sum_k{\cal A}_ku_{\phi_{(1)}}(\XX)+\cc
	\label{potencjal1}
	\end{eqnarray}
of modes
	\begin{eqnarray}
u_{\phi_{(1)}}(\XX)=\frac{1}{\sqrt{2\pi}}\frac{1}{(\omega\eta)^2}\left[1+\frac{1}{{\tt i}\omega\eta}\right]
\exp\left[{{\tt i}(\KX-\omega\eta)}\right].\nonumber\\
	\label{eq:padajaca}
	\end{eqnarray}
In the matter dominated universe equation (\ref{eq:PertDust}) 
is solved by an arbitrary combination of the constant or decreasing modes
	\begin{eqnarray}
\phi_{(2)}(\XX)=
\sum_k\left[\alpha_k-\frac{\beta_k}{5(\eta+\es)^5}\right]\exp\left[{{\tt i}\KX}\right]+\cc
	\end{eqnarray}
To assure the continuity of the first and the second fundamental forms at the decoupling hypersurface the
coefficients ${\cal A}_k$, $\alpha_k$ and $\beta_k$ must be related by
	\begin{eqnarray}
\alpha_k&=&-\frac{1}{5\sqrt{2\pi}}\left[\frac{1}{(\omega\es)^2}+
{\tt i}\left(\frac{2}{\omega\es}-\frac{1}{(\omega\es)^3}\right)\right]{\cal A}_k\exp\left[-{\tt i}\omega\es\right]\label{alfa}, \\
\beta_k&=&-\frac{64}{\sqrt{2\pi}}\es^5\left[\frac{3}{(\omega\es)^2}+
{\tt i}\left(\frac{1}{\omega\es}-\frac{3}{(\omega\es)^3}\right)\right]{\cal A}_k\exp\left[-{\tt i}\omega\es\right].
\label{beta}
	\end{eqnarray}
Then the generic solution after decoupling is
	\begin{eqnarray}
\phi_{(2)}=\sum_k{\cal A}_ku_{\phi_{(2)}}(\XX)+\cc
	\label{potencjal2}
	\end{eqnarray}
where 
	\begin{eqnarray}
u_{\phi_{(2)}}&=&\frac{1}{5\sqrt{2\pi}}
\left\{
\frac{1}{(\omega\es)^2}\left[
6\left(\frac{2}{1+\eta/\es}\right)^5-1\right]
+{\tt i}\left[
\frac{2}{\omega\es}\left[
\left(\frac{2}{1+\eta/\es}\right)^5-1\right]\right.\right.\nonumber\\
&&+\left.\left.\frac{1}{(\omega\es)^3}\left[
1-6\left(\frac{2}{1+\eta/\es}\right)^5\right]
\right]
\right\}\exp\left[{{\tt i}(\KX-\omega\es)}\right].
	\label{eq:zszyta}
	\end{eqnarray}
The pair of solutions (\ref{eq:padajaca}) and (\ref{eq:zszyta}) combines into a~single 
mode $\{u_{\phi_{(1)}},u_{\phi_{(2)}}\}$ regular everywhere in the space-time. 
The density contrast and the four-velocity corrections read
	\begin{eqnarray}
\frac{\delta\epsilon}{\epsilon}=2
	\left[
-\pez+\frac13\left(\frac{\apeta}{\aeta}\right)^{-2}
\Delta\phi-\left(\frac{\apeta}{\aeta}\right)^{-1}\pp{\pez}{\eta}
	\right]
	\label{eq:delta}
	\end{eqnarray}
	\begin{eqnarray}
\delta\v4^{\mu}=
\left(\pez,\frac{2}{3}\apeta^{-2}
\frac{\partial}{\partial\eta}\left(\aeta
\nabla
\pez\right)\right)
	\label{eq:velo}
	\end{eqnarray}
with $\pez$ given by (\ref{potencjal1})--(\ref{eq:padajaca}) or (\ref{potencjal2})--(\ref{eq:zszyta}) respectively. 
The freedom to choose ${\cal A}_k$ is the only freedom left. Initial data on $\Sigma_d$ determined by (\ref{alfa})--(\ref{beta}) are consistent with both equations, (\ref{eq:PertRad}) and (\ref{eq:PertDust}). While evolved forward or backward in time they restore the perturbation field in each of considered epochs.

\section{The transfer function}

We introduce dimensionless variables
$\ka$, $\iks$ and $\te$ 
based on the characteristic time-scale $\es$ (time of decoupling)
and a characteristic length-scale ~$\es/\sqrt3$ (length of the sound horizon)
	\begin{eqnarray}
\ka=\frac{\es}{\sqrt3}k,\quad 
\iks=\frac{x}{{\es}\slash{\sqrt3}},\quad \te=\frac{\eta}{\es}.
	\end{eqnarray}
$\ka$ and $\te$ relate to observables
        \begin{eqnarray}
\te&=&\frac{2\sqrt{1+z_d}}{\sqrt{1+z}}-1\\
\ka&=&\left(\frac{4\pi^3}{3\sqrt{3}G H_0 {\cal M}}\right)^{1/3} (1+z_d)^{-1/2}
	\end{eqnarray}
where $H_0$ stands for the Hubble constant, $z_d$ is the redshift of decoupling, ${\cal M}$ is the mass of sphere 
of the wavelength diameter.

The density perturbations expressed in dimensionless variables take the form
		\begin{eqnarray}
\delta_{(1)}=\sum_k{\cal A}_k u_{\delta_{(1)}}(\YX)+\cc\label{d1}\\
\delta_{(2)}=\sum_k{\cal A}_k u_{\delta_{(2)}}(\YX)+\cc\label{d2}
		\end{eqnarray}
where
\def\kuku{2 \todwr\left(3+\ik(3+\ik)\right)}
\def\lala{\left[1+\ik(1+2\ik)-\kuku\right]}
	\begin{eqnarray}
u_{\delta_{(1)}}&=&\frac{-2\ikate}{(\ka \te)^4}
(2+ \ikate(2+ \ikate(2+ \ikate)))
\exp({\tt i}\kappa (\iks-\te))
	\label{umon1} 
	\end{eqnarray}
and
	\begin{eqnarray}
u_{\delta_{(2)}}&=&-\frac{\ik}{\ka^4}\exp(\ik(\iks-1))
\left[ \kuku\right. -\frac{2}{5}\left[\ka^2\polowka^2+1\right]\nonumber\\
&&{\times}\left.\lala\right].
	\label{umon2} 
	\end{eqnarray}
The linear dispersion relation $\omega=k/\sqrt{3}$ is fulfilled. Perturbations of both the energy density and the expansion rate are continuous at the transition. 

The transfer function~\cite{Padnamabhan2006} normalized to the decoupling epoch $\tau=1$ reads
\def\udd{u_{\delta_{(2)}}(\YX)}
\def\uconj{u^{\ast}_{\delta_{(2)}}(\YX)}
\def\licznik{{\sqrt{\udd \uconj}}}
\def\pu{u_{\delta_{(2)}}(1,\iks)}
\def\puconj{u^{\ast}_{\delta_{(2)}}(1,\iks)}
\def\mianownik{{\sqrt{\pu \puconj}}}
	\begin{equation}
T(\ka, \te) = \frac{\licznik}{\mianownik}.
\end{equation}
On the strength of (\ref{umon2}) one  has
	\begin{eqnarray}
T^2(\ka, \te)&=&\frac{1}{\gr800}\left[{\dziewiec\TCa^2}
-2\trzy\TCb\TCa +\jeden\TCb^2\right]\nonumber\\
\TCa&=&\Ta\nonumber\\
\TCb&=&\Tb
	\label{transfer}
	\end{eqnarray}
and therefore,
\begin{equation}
a_k(\te)a^{\ast}_k(\te)=T^2(\ka, \te)  a_k(1)a^{\ast}_k(1)
\end{equation}
with the usual definition of $a_k$ given by
\begin{eqnarray}
\delta_{(2)}(\YX)&=&\sum_k{\cal A}_k u_{\delta_{(2)}}(\YX)+\cc\nonumber\\
            &=&\sum_k{a}_k(\te) \exp({\tt i} \ka \iks)+\cc
		\end{eqnarray}
The transfer function (\ref{transfer}) differentiate perturbations of various length scales. The contour plot of $T({\cal M},{z})$ as the function of mass and redshift (in the logarithmic scale) is shown on Figure~1. The mass $\cal M$ is expressed in the solar mass units, the decoupling redshift $z_d=1200$ is assumed. Conditions (\ref{alfa}) and (\ref{beta}) do not restrain the perturbations spectrum. The invariant spectrum $A_k A^\ast_k$ (or equivalently the coefficients $a_k(t_d)$ on $\Sigma_d$) remains a~free parameter of the model.

\begin{figure}[th]
\centerline{\includegraphics[width=.45\textwidth]{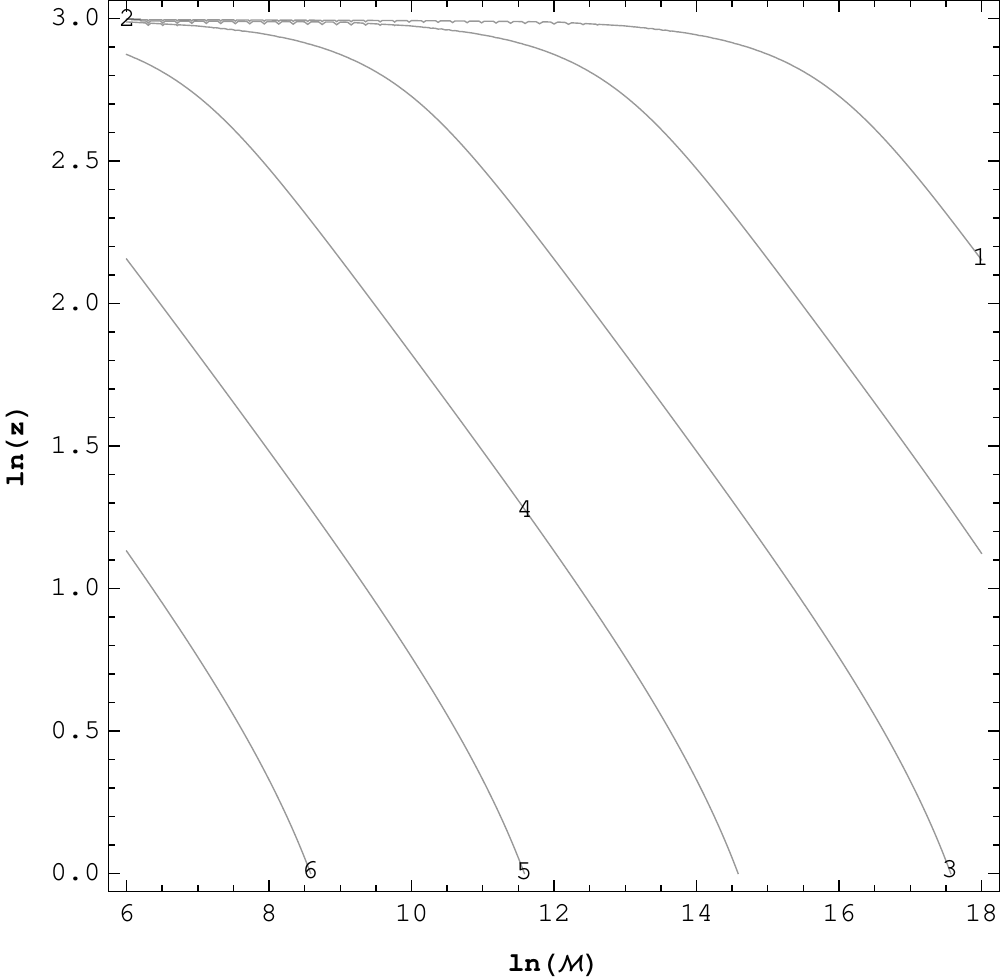}}
\vspace{-3mm}
\caption{Logarithmic contour map of the transfer function   $\ln T({\ln \cal M},\ln z)$. 
$z$ stands for redshift,  ${\cal M}$  is the perturbation mass expressed in the Solar mass units. }
\end{figure}


\end{document}